\documentclass[apj,twocolumns]{emulateapj}

\shorttitle{Radiative cluster winds}
\shortauthors{Ary Rodr\'\i guez-Gonz\'alez et al.}

\begin{document}

\title{The formation of filamentary structures in radiative cluster winds}

\author{A. Rodr\'\i guez-Gonz\'alez\altaffilmark{1}
, A. Esquivel\altaffilmark{1} , A. C. Raga\altaffilmark{1},
J. Cant\'o\altaffilmark{2}}
\altaffiltext{1}{Instituto de Ciencias Nucleares, Universidad
Nacional Aut\'onoma de M\'exico, Ap. 70-543,
04510 D.F., M\'exico}
\altaffiltext{2}{Instituto de Astronom\'{\i}a, Universidad
Nacional Aut\'onoma de M\'exico, Ap. 70-264,
04510 D.F., M\'exico}

\email{ary, esquivel, raga@nucleares.unam.mx}

\begin{abstract}
We explore the dynamics of a ``cluster wind'' flow in the
regime in which the shocks resulting from the interaction of winds
from nearby stars are radiative. We first show that for a cluster
with T~Tauri stars and/or Herbig Ae/Be stars, the wind interactions
are indeed likely to be radiative. We then compute a set of four,
three dimensional, radiative simulations of a cluster of 75
young stars, exploring
the effects of varying the wind parameters and the density of
the initial ISM that permeates the volume of the cluster. These
simulations show that the ISM is compressed by the action
of the winds into a structure of dense knots and filaments.
The structures that are produced resemble in a qualitative
way the observations of the IRAS 18511+0146 of Vig et al. (2007).
\end{abstract}


\section{Introduction}

In a recent paper, Vig et al. (2007) reported JCMT-SCUBA and Spitzer
(IRAC \& MIPS) observations of the region around
IRAS 18511+0146, and deduced that there is a young cluster of Herbig Ae/Be
stars (probably also having lower mass stars), which is still embedded
in a dense, highly extinguished region. The interstellar medium (ISM)
within this cluster has
a complex structure of dense clumps and filaments, which are seen as
 structured extinction of the background infrared (IR) emission.
In the present paper, we explore whether or not this clump/filament
structure could be the result of the stirring up of the ISM by the winds
from the young stars.

This cluster appears to be
a younger, more embedded version of clusters of Herbig Ae/Be stars such
as the ones described by Testi et al. (1997, 1999). These clusters
typically have less than 100~stars within regions of $\sim 1$~pc
size (corresponding to $D\sim 0.1$~pc separations between the cluster stars).

The region between the stars in this kind of cluster will of course
be stirred up by the massive winds ejected by the young stars. These
winds will have high, ${\dot M}_w>10^{-7}$~M$_\odot$yr$^{-1}$
mass loss rates and terminal velocities $v_w=100-400$~km~s$^{-1}$.
These parameters, together with the $D\sim 0.1$~pc separations between
the cluster stars (see above) imply that the shock interactions between
the winds from nearby stars might be radiative.

The ``cluster wind'' resulting from the combined effect of all of the
stellar winds will in this case be very different from the non-radiative
cluster winds studied by Chevalier \& Clegg (1985) and Cant\'o et
al. (2000). These papers have all studied cluster 
winds with no radiative losses (Chevalier \& Clegg 1985, Cant\'o et
al. 2000, Raga et al. 2001 and Rockefeller et al. 2005,
Rodr\'{i}guez-Gonz\'alez et al. 2007), or with relatively small 
radiative losses (Silich et al. 2004, Tenorio-Tagle et al. 2005), and
are applicable for clusters of massive 
stars (with wind velocities of $\sim 1000$~km~s$^{-1}$ which result
in non-radiative wind-wind interactions).

Motivated by the observations of IRAS 18511+0146 of Vig et al. (2007),
we have studied the flow resulting from a cluster of stars with wind
interactions that produce radiative shocks. As this problem
has not been studied before, we present a series of idealized
models in which all of the stars in the cluster have identical,
isotropic winds, and in which the initial ISM permeating the volume
of the cluster is homogeneous. Because of these simplifications, our
models should be regarded as an exploration of the properties of
highly radiative cluster wind flows rather than an attempt to model
a particular object (namely, IRAS 18511+0146).

In the present paper, we first study for which combinations of
parameters (mass loss rate ${\dot M}_w$, stellar wind velocity $v_w$,
and separation $D$ between nearby cluster stars) one 
obtains a highly radiative cluster wind flow (\S 2). We then compute
four 3D simulations of cluster winds in the highly radiative regime (\S 3),
and obtain predictions of the highly structured column density maps
that result from our simulation (\S 4).
Finally, in \S 5 we present our conclusions.

\section{Radiative losses in a cluster wind}

Let us consider a stellar cluster with a local stellar density $n$ (=number
of stars per unit volume), of stars with identical, isotropic winds
with a mass loss rate ${\dot M}_w$ and a terminal wind velocity $v_w$.
The typical separation between stars then is $D=n^{-1/3}$.

The two-wind shock interactions between nearby stars occur at a typical
distance $\sim D/2$ from each of the stars, so that the typical
pre-shock densities have values
\begin{equation}
n_{pre}={{\dot M}_w\over 1.3 m_H\pi D^2 v_w}\,,
\label{npre}
\end{equation}
where $m_H$ is the H mass, and we have assumed a 90\%\ H and
10\%\ He particle abundance.

The shock interactions between nearby stars will be radiative if
the cooling distance $d_{cool}$ satisfies the condition
\begin{equation}
\kappa \equiv {d_{cool}\over D}<1\,.
\label{k}
\end{equation}

In order to estimate the cooling distance, we proceed as follows. We
consider the head of the stellar bow shock structures that will be formed
with a shock velocity of  $\approx v_w$ (i. e., equal to the stellar
wind velocity) and a preshock density given by equation (\ref{npre}).
We then use the fact that the cooling distance (for shocks with cooling
in the low density regime) scales as the inverse of the pre-shock density
to write
\begin{equation}
d_{cool}(n_{pre},v)=\left({100\,{\rm cm^{-3}}\over n_{pre}}\right)
d_{c,100}(v)\,,
\label{dc}
\end{equation}
where $v$ ($=v_w$) is the shock velocity, and $d_{c,100}(v)$ is the
cooling distance behind a shock with $n_{pre}=100$~cm$^{-3}$. We
then estimate the function $d_{c,100}(v)$ by carrying out a fit to
the cooling distance (to $10^4$~K) obtained from the ``self-consistent
preionization'', steady, plane-parallel models of Hartigan, Raymond
\& Hartmann (1987). We use
\begin{displaymath}
d_{c,100}(v)={3.3\times 10^{14}\,{\rm cm}}\,
\left[1+\left({{\rm 135\,km\,s^{-1}}\over v}\right)^{10.7}\right]\times\\
\end{displaymath}
\begin{equation}
~~\left\{1-\exp\left[-\left({v\over {\rm 200\,km\,s^{-1}}}\right)^6\right]
\right\}
\left({v\over {\rm 100\,km\,s^{-1}}}\right)^{4.2}\,,
\label{dcfit}
\end{equation}
which fits the computed cooling distances with a relative accuracy
of better than
15\% in the $v=100\to 400$~km~s$^{-1}$ shock velocity range. We
note that large deviations between this fit and the values of Hartigan
et al. (1987) occur for $v<100$~km~s$^{-1}$, so that the fit should not
be applied for lower shock velocities.

Combining equations (\ref{npre}-\ref{dcfit}) we then obtain
\begin{equation}
\kappa={d_{cool}\over D}={1.1\times 10^{-2}}
\left({{\rm 10^{-6} M_\odot yr^{-1}}\over {\dot M}_w}\right)
\left({D\over {\rm 0.1 pc}}\right)\,f(v_w)\,,
\label{kfit}
\end{equation}
where
\begin{displaymath}
f(v)=\left[1+\left({{\rm 135\,km\,s^{-1}}\over v}\right)^{10.7}\right]\times
\end{displaymath}
\begin{equation}
\left\{1-\exp\left[-\left({v\over {\rm 200\,km\,s^{-1}}}\right)^6\right]
\right\}
\left({v\over {\rm 100\,km\,s^{-1}}}\right)^{5.2}\,.
\label{f}
\end{equation}

\begin{figure}
\centering
\includegraphics[width=8.5cm]{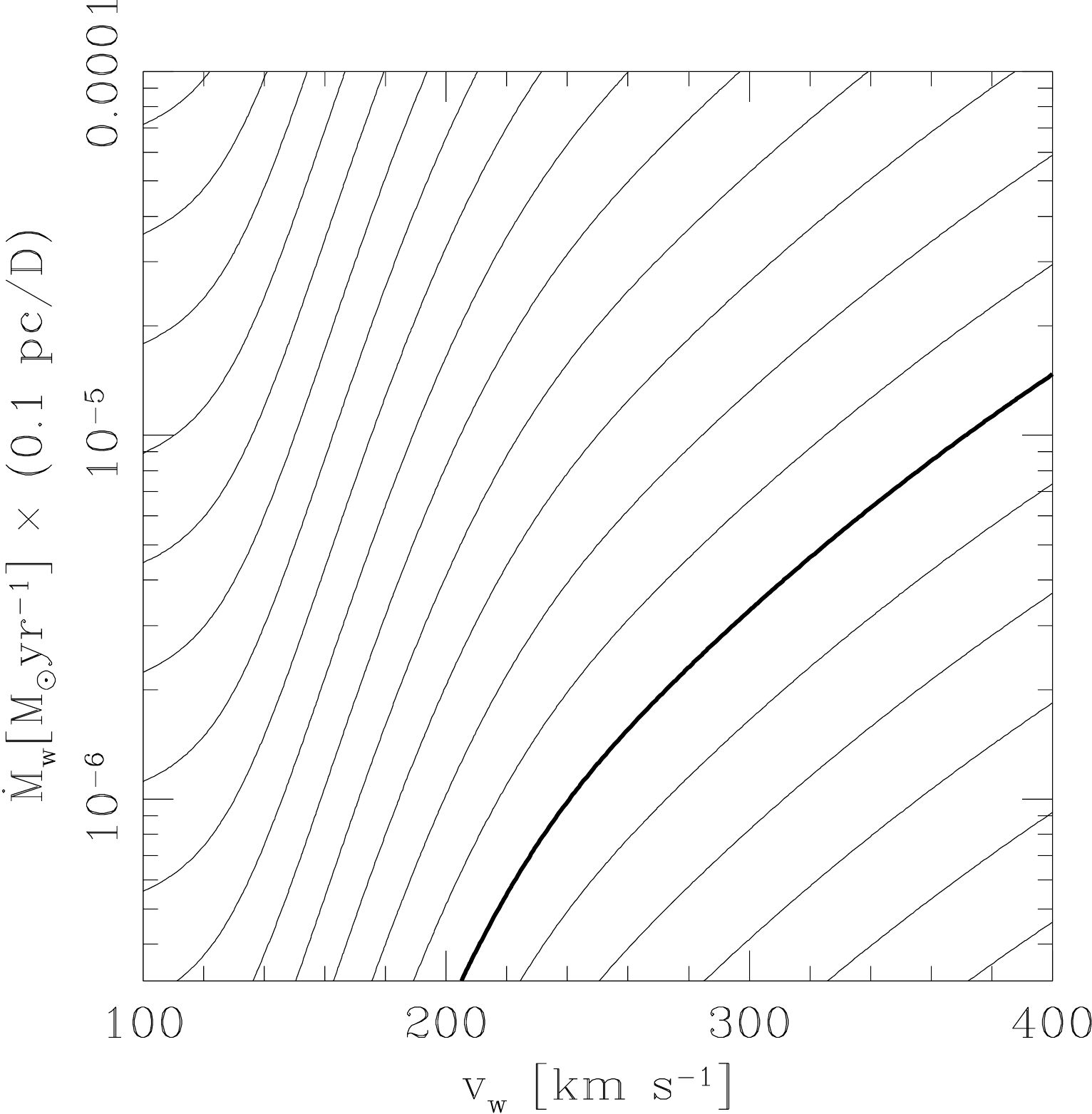}
\caption{This diagram shows the ``cooling parameter'' $\kappa=
d_{cool}/D$ (where $D$ is the typical the separation between stars) as
a function of ${\dot M}_w \times (0.1$pc$/D)$ and $v_w$,
where ${\dot M}_w$ is the mass loss rate and $v_w$ the wind
velocity. The thick curve shows the $\kappa=1$ contour, the
other contours represent values of $\kappa$ at successive
factors of two below (top left region of the diagram) or above (bottom right
region of the diagram) unity. The region above the thick curve therefore
represents the parameter space in which the two-wind interactions in
a cluster wind flow are radiative.}
\label{cool}
\end{figure}

In Figure 1, we plot the cooling parameter $\kappa$ (see equation
\ref{kfit}) as a function of $v_w$ and ${\dot M}_w/D$.
This figure shows that a substantial part of the
parameter range that would be expected for a cluster of low or intermediate
mass stars produces cooling parameters $\kappa<1$.

A cluster wind in this ``high cooling regime'' will have dense, cool
structures resulting from the radiative shocks in the interactions
between nearby stars. In the following section, we present a numerical
simulation of a flow in this regime.

\section{A 3D simulation of a radiative cluster wind}

\subsection{The numerical setup}

We have carried out 4 numerical simulations in which we place
75 stars which are uniformly distributed inside a sphere of
radius $R_c=0.5$~pc. The stars have identical mass loss rates
${\dot M}_w=10^{-6}$ (models M1 and M2), $4\times 10^{-6}$ (M3)
or $10^{-5}$~M$_\odot$yr$^{-1}$ (M4) and wind velocities
$v_w=100$ (M1), 200 (M2 and M3) or 300~km~s$^{-1}$ (M4). These
parameters are listed in Table 1. These wind parameters span
a parameter range which is appropriate for low and intermediate
mass stars, and have been chosen so as to produce a cluster wind
in the highly radiative regime (see below).

With the number of stars and the volume of the cluster, we
obtain a density of stars $n=133.7$~pc$^{-3}$, thus
a typical separation between stars of $D=n^{-1/3}=0.20$~pc.
 For these values for $D$, ${\dot M}_w$ and $v_w$,
using equations (\ref{kfit}-\ref{f}) we obtain
$\kappa=0.008$, 0.518, 0.130 and 0.666 for models M1-M4,
respectively. In other words, the post-wind bow
shock cooling distances have values $\sim 0.5\to 30$~\%\ of the typical
separations between stars, placing the simulations in the highly
radiative regime.

The stars are placed randomly within the volume of the spherical
clusters, elliminating the positions that appear within a distance
of $<0.035$~pc from any of the previously generated stellar positions.
We then impose the isotropic stellar wind condition within spheres
of radii $R_w=0.0172$~pc centered at each of the stellar positions. A
uniform, $T_w=5000$~K temperature is imposed within each of the wind
sources.  The rest of the computational domain is filled with
a homogeneous medium with number density $n_{env}=10^4$ (models
M1 and M2) or $5\times 10^4$~cm$^{-3}$ (models M3 and M4, see Table 1)
and a temperature $T_{env}=100$~K. The stellar winds are
``turned on'' simultaneously at the beginning of the time-integration.
The winds and the environment 
have neutral H, and a seed electron density which is assumed
to come from singly ionised Carbon.

The gasdynamic equations, together with a rate equation
for neutral Hydrogen are integrated using
the ``yguaz\'u-a'' code (Raga et al. 2000, 2002). The energy equation
includes the cooling function described by Raga \& Reipurth
(2004). This cooling function is appropriate for describing the
cooling of the shocked wind material. However, it is not necessarily
appropriate for the shocked ISM, which is likely to be initially
molecular. At the relatively low resolution of our simulations,
however, the details of the cooling function are unlikely to be
important, as the cooling regions are at best marginally resolved.

The $T_{env}=100$~K chosen for the initial configuration of the environment
is higher than the $\sim 10$~K temperature expected for a molecular cloud.
This artificially high choice of temperature is consistent with the fact
that our cooling function (see above) is appropriate only for partially
ionized gas, as it does not include the cooling processes important at
temperatures $\sim 10$~K.

The numerical integrations are done in a
cubic domain with a size of $1.2$~pc, using a
5-level, binary adaptive grid with a maximum resolution
of $512$ (M1) or 256 (M2-M4) points along each of the three axes.
The maximum resolution level is only allowed within the spheres
in which the stellar wind conditions are imposed.
Outflow conditions are imposed on all of  the grid
boundaries.

Together with the gasdynamic equations and the rate equation for
neutral H (see above), we have integrated an advection equation for
a passive scalar. This scalar has a positive value for the environmental
material, and a negative value for the stellar winds. In this way, we
can at all times trace which regions of the computational domain are
occupied by wind or by environmental material (identified by the
sign of the passive scalar).

\subsection{Model results}

We have carried out time integrations from $t=0$ (when the stellar
winds are ``turned on'') up to $t=1.9\times 10^5$~yr for models
M1-M4. Figure 2 shows a ``volume-rendered'' depiction of the
3D density structure of model M1 at this
final time. We see that there is a dense shell (which has partially
left the computational domain) which is being pushed out by the stellar
winds into the surrounding environment. This shell is formed by
material from the winds and by part of the
ISM material which was initially permeating the volume of the cluster.
Also, we see that there is a complex structure of filaments and
clumps, which are formed from material that was initially present
in intercluster medium and wind material that has gone through
the stellar wind bow shocks and cooled to low temperatures (and high
densities).
\begin{table}[!h]
\centering
\begin{minipage}{80mm}
\caption{Model parameters}
\begin{tabular}{@{}llrrrrrr@{}}
\hline
Model&$v_w$& $\dot{M}_w$& n$_{cloud}$ & $\kappa$%
\footnote{Using the fit to the models of Hartigan et al. (1987),
 i.e. equation~(\ref{kfit})} &resolution \\

          &km s$^{-1}$&M$_\odot$ yr$^{-1}$&cm$^{-3}$ &  &  pixels\\
\hline
\hline
M1       &    100   &    10$^{-6}$        &    10$^{4}$       &   0.004 &
512$^3$ \\
M2       &    200   &    10$^{-6}$        &    10$^{4}$       &   0.259 &
256$^3$ \\
M3       &    200   & 4$\times$10$^{-6}$  & 5$\times$10$^{4}$ &   0.065 &
256$^3$ \\
M4       &    300   &    10$^{-5}$        & 5$\times$10$^{4}$ &   0.333 &
256$^3$ \\
\hline
\end{tabular}
\end{minipage}
\end{table}

\begin{figure*}
\centering
\includegraphics[width=15cm]{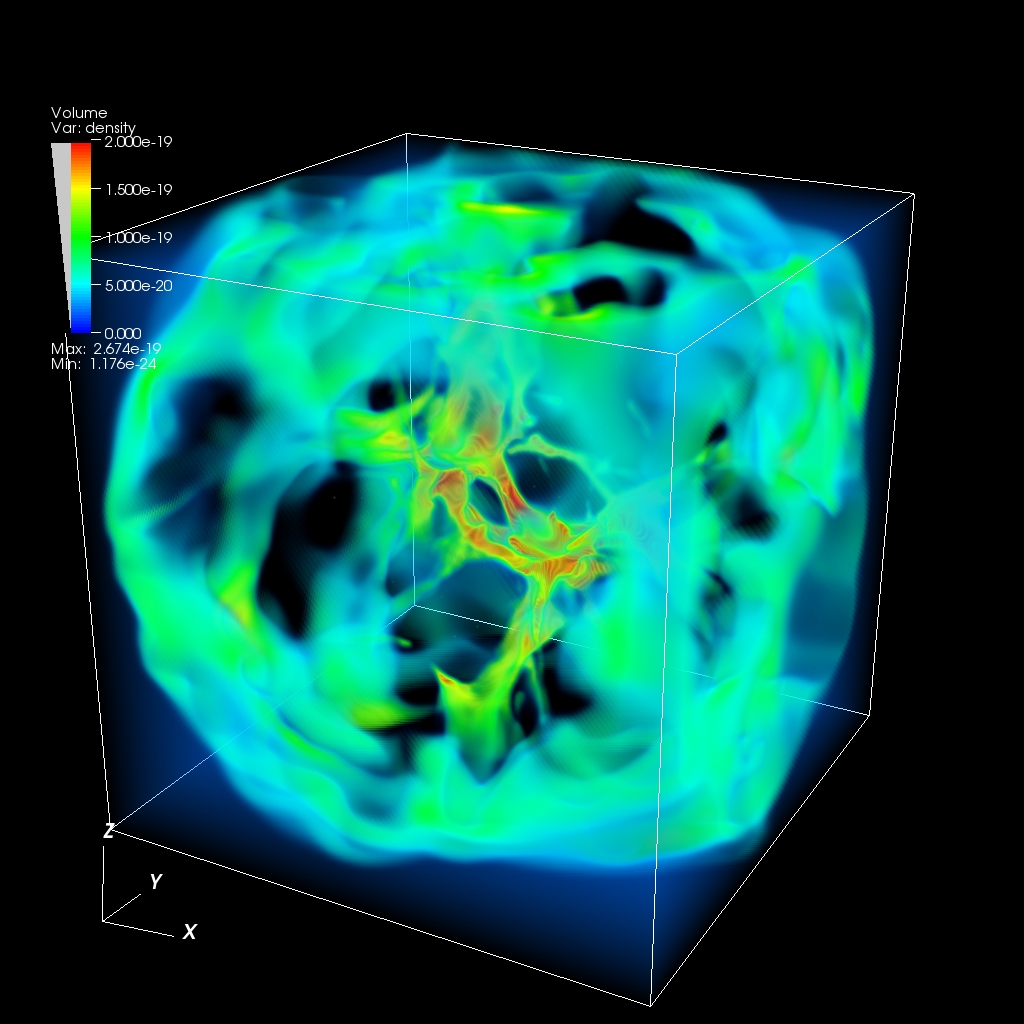}
\caption{Volume rendition of the density structure of model M1
  after a $t=1.9\times 10^5$~yr time-integration. The grey-scale 
(color-scale in the online version) shows the
  density values in g cm$^{-3}$, linearly scaled, with an opacity that is also
  linearly proportional to the density. The cubic domain has
  a size of 1.2 pc.}
\label{3d}
\end{figure*}

In Figure 3, we show column density maps (obtained by
integrating the number density along the $z$-axis of the
computational domain) at integration times $t=19$, 3.2,
11 and $3.2\times 10^4$~yr for models M1, M2, M3 and M4,
respectively. These times have been chosen so that the
outer, dense shell is starting to leave the computational
domain in each of the models. All of the maps have a
central condensation, which has a peak column density of
9.11, 12.8, 66.6 and $36.3\times 10^{22}$~cm$^{-2}$ for
models M1-M4, respectively.

Figure 4 shows column density maps for models M1-M4 computed
only with the material from the ISM which originally permeated
the computational domain. In these maps, the central condensation
has a peak column density of 1.82, 10.8, 29.7 and
$33.6\times 10^{22}$~cm$^{-2}$ for models M1-M4, respectively.
Comparing the peak column densities for ISM material only (Figure 4)
and the column densities for ISM+wind material (Figure 3), we see
that while for model M1 the main contribution to the column density
of the central condensation
comes from the wind material, for models M2-M4 the main contribution
to this column density comes from the material in the initial ISM.

We have also computed the fraction $f_m$ of mass of the initial
ISM which is still present within the volume of the cluster
(i. e., within a sphere of radius $R_c=0.5$~pc) for the time
frames shown in Figure 4. We obtain $f_m=0.076$, 0.026, 0.26 and
0.052 for models M1-M4, respectively. Therefore, for model
M3, almost 75\%\ of the initial ISM mass has already been expelled
from the volume of the cluster (most of this mass being present
in the expanding shell pushed out by the cluster wind flow), and
for the other three models more than 90~\%\ of the ISM mass has been
expelled.

As the material from the winds is unlikely to have dust, the
column density distributions of the ISM material (see Figure 4)
are proportional to the expected extinction. If we use a standard
conversion factor $A_V=10^{-21}N_H$, where $A_V$ is the optical
extinction and $N_H$ is the hydrogen column density
in cm$^{-2}$, the peak values of $A_V$ corresponding to
the central condensations are $A_V\approx 18$, 110, 300 and 340, for
models M1-M4, respectively.

Finally, in Figure 5 we present a series of column density maps
(computed only for the material of the initial ISM)
illustrating the time-evolution obtained from model M3. From this
time-sequence, we see that we can already identify the structure
that forms the central condensation at  $t=1.6\times 10^4$~yr. 
This condensation persists for $\sim 10^5$~yr
(i. e., to the end of the time-integration). Several other features
can be seen in both the $t=4.8\times 10^4$ and
$t=7.9\times 10^4$~yr frames, indicating that the filaments have
lifetimes $\sim 10^4$~yr.

\begin{figure*}
\centering
\includegraphics[width=15cm]{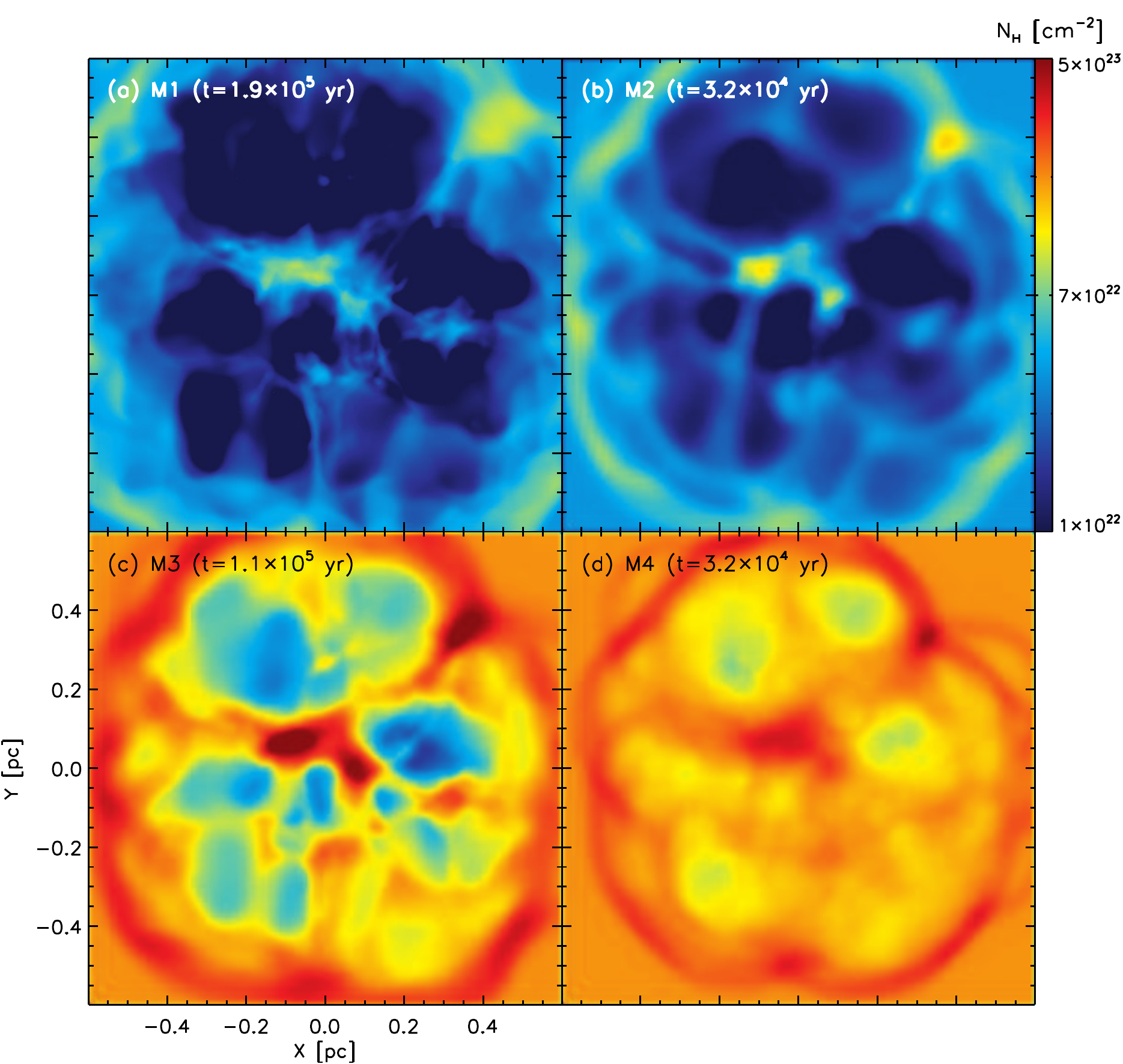}
\caption{Column density maps obtained from models M1-M4 by integrating
the number density along the $z$-axis. 
The stratifications are shown with the logarithmic gray-scale
(color-scale in the online version) given by the bar on the top right
plot. The integration time for each model is indicated in
the label at the top of each panel.}
\label{av1}
\end{figure*}

\begin{figure*}
\centering
\includegraphics[width=15cm]{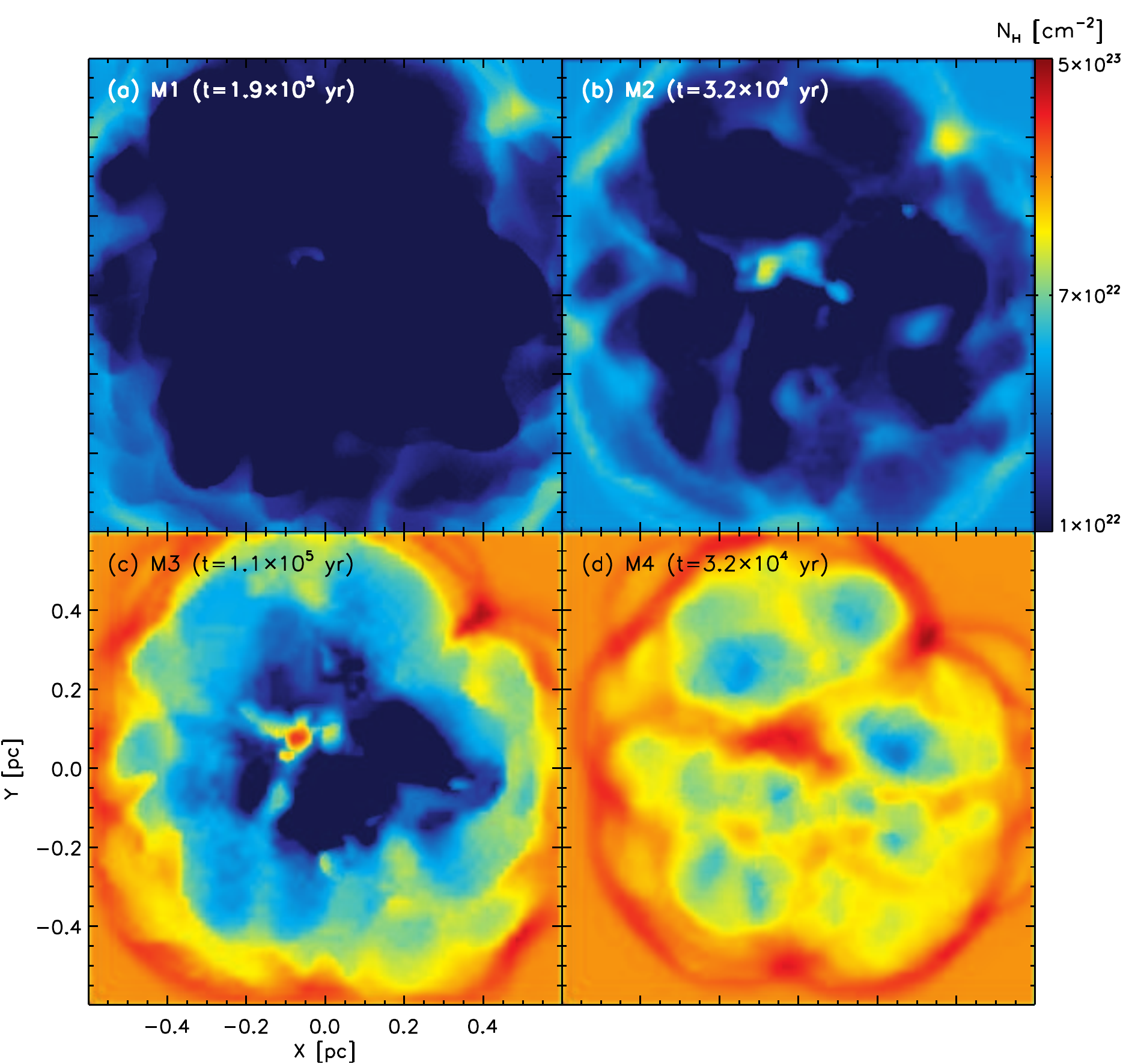}
\caption{ISM column density maps obtained from models M1-M4 by integrating
the number density of the material in the initial ISM along the $z$-axis. 
The stratifications are shown with the logarithmic gray-scale
(color-scale in the online version) given by the bar on the top right
plot. The integration time for each model is indicated in
the label at the top of each panel.
}
\label{av2}
\end{figure*}

\begin{figure*}
\centering
\includegraphics[width=15cm]{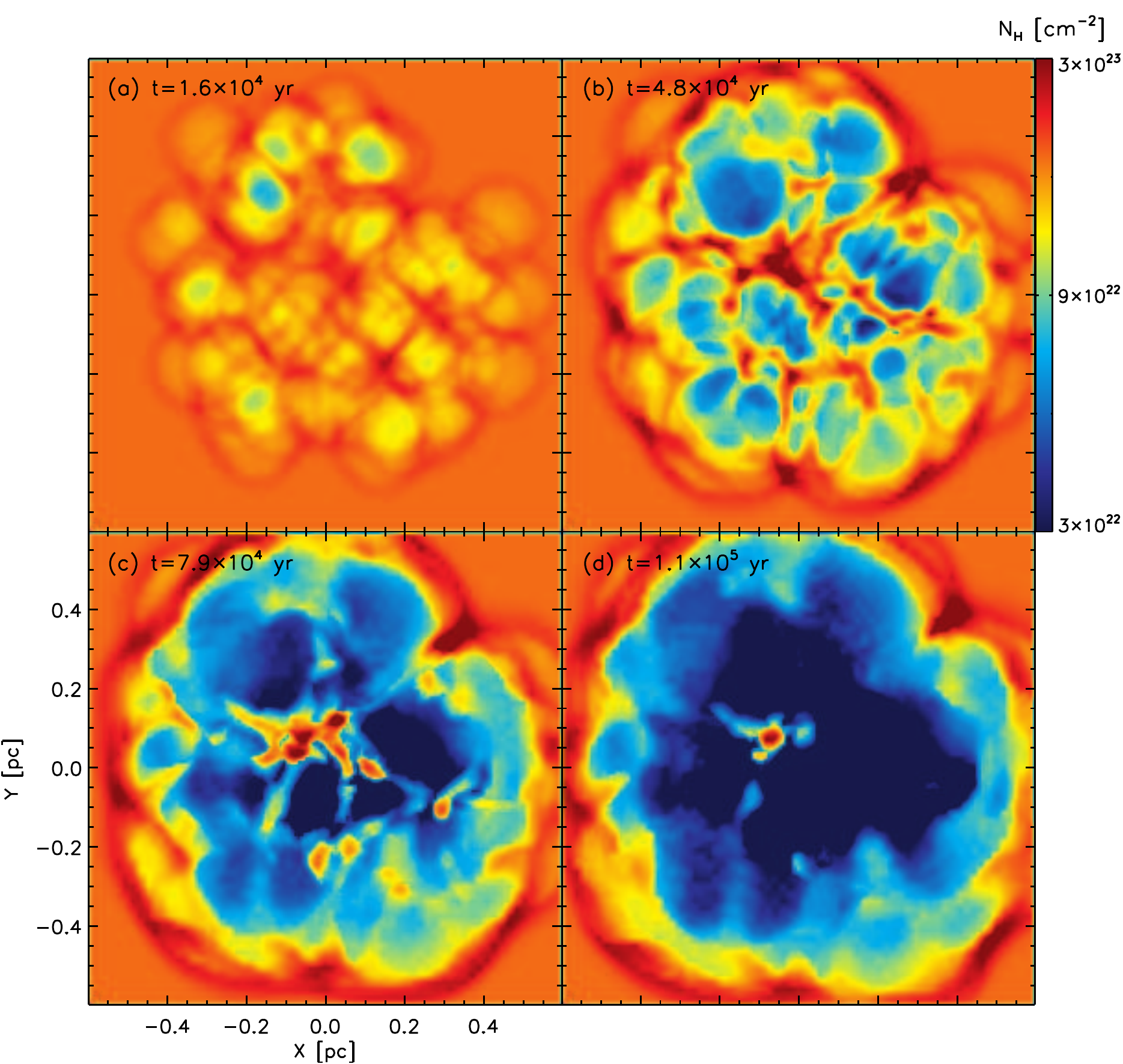}
\caption{Time-sequence showing the evolution of the ISM
column density structure of model M3 (obtained by integrating the density
of the material in the initial ISM along the $z$-axis).
The stratifications are shown with the logarithmic gray-scale
(color-scale in the online version) given by the bar on the top right
plot.}
\end{figure*}

\section{Conclusions}

Recent Spitzer observations of a young cluster of Herbig Ae/Be stars
(Vig et al. 2007) show that the IR extinction towards IRAS 18511+0146
has a curious structure of filaments and clumps. This observation
has motivated us to study the formation of dense, neutral structures
as a result of the interaction between the winds of the young stars.

We find that if the Herbig Ae/Be stars have wind velocities $v_w<
200$~km~s$^{-1}$, the stellar wind interactions between nearby
stars are highly radiative for mass loss rates as low as
${\dot M}_w \approx 3\times 10^{-7}$M$_\odot$yr$^{-1}$
(see Figure 1 and equation \ref{kfit}). However, for higher
wind velocities, considerably higher mass loss rates
are required for the wind interactions to be radiative.
For example, for $v_w=400$~km~s$^{-1}$, mass loss rates
${\dot M}_w > 10^{-5}$M$_\odot$yr$^{-1}$ are required
(see Figure 1). The values quoted in this paragraph are for
an average separation of 0.1~pc between the stars in the cluster.

These values indicate that in a cluster of Herbig Ae/Be stars
many of the wind interactions between nearby stars are likely
to be in the highly radiative regime. We then compute a 3D
simulation of a cluster wind flow in this regime, and we show that
it does lead to the production of a structure of dense filaments and
clumps.

From our simulations, we obtain predicted column density maps which
show spatial structures which are qualitatively similar to the
observations of IRAS 18511+0146 (Vig et al. 2007). Both the
predicted and the observed maps show a structure of dense clumps
and filaments, distributed within as well as outside the volume
of the clump. We find that in all simulations a central clump
is produced, which has a lifetime of $\sim 10^5$~yr.

It is more difficult to carry out a more quantitative comparison
between our predictions and the observations of IRAS 18511+0146.
Vig et al. (2007) first suggest that the filament/clump structures
in this object have $A_V\sim 6$-$8$ extinctions (deduced from their
Spitzer IR observations). However, when they compare these values with
the extinction deduced from millimetre observations, they conclude
that the IR maps probably have a strong contribution from foreground
emission, and that the real values of the extinction should be
$A_V\sim 50$. Therefore, only our model M1 produces
visual extinctions which are too low compared with the
ones deduced from the observations.

We end by noting that the simulation presented in this paper is
the first attempt that has been made for modelling the interaction
of the outflows from a cluster of young, low and intermediate mass
stars. Our simulation was done assuming that~:
\begin{itemize}
\item the stellar winds are ``turned on'' simultaneously,
\item the region is initially filled by a homogeneous medium,
\item the stellar outflows are isotropic,
\item all of the stellar winds are identical.
\end{itemize}
It is particularly important to remove this last approximation if one wants
to model IRAS 18511+0146 in detail, as this cluster harbours the massive
protostar IRAS 18511 A.

The other assumptions in our models (listed above)
are also not realistic, and the effects of removing them should be
explored in future work. For example, assuming that the winds are isotropic
is somewhat dubious in the context of a cluster of young stars. The more
massive stars in such a cluster will evolve faster from an early, collimated
outflow phase into a later phase in which a more isotropic wind is produced.
The lower mass stars in the cluster will evolve less rapidly, and will remain
in a collimated outflow phase for a longer period. For example, for a cluster
with an age of $\sim 10^5$~yr one might expect to have a few more massive
($\sim 10$M$_\odot$) stars with isotropic winds, and a larger number of lower
mass stars still producing jets. This combination of isotropic and collimated
winds will result in flow configurations that might be applicable when
more detailed observations of the wind interaction in clusters of young stars
become available.

Another point that would be worthwhile to explore is the effect of the
cooling function. As we have described in \S 2, our simulations are
computed with a parametrized cooling function appropriate for partially
(or fully) ionized gas. If one included a description of the non-equilibrium
chemistry, and used it to compute a molecular
cooling function, one would obtain cooling to lower temperatures in the
dense filaments of compressed cloud material. This would lead to the
production of narrower and denser filaments. A calculation including these
effects should also have a substantially higher resolution in order to
be able to resolve the associated cooling regions.

The conclusion that we obtain from the work described above is that the
filamentary structure observed in IRAS 18511+0146 by Vig et al. (2007)
might be the result of the surrounding ISM being compressed into dense
sheets by the wind interactions between the cluster stars. However,
it is of course clear that at least part of the observed structures could have
been present initially in the dense cloud from which the cluster stars
were formed.

\begin{acknowledgements}
We thank the anonymous referee for very relevant comments that
resulted in a substantial revision of the original version
of this paper. We acknowledge support from
the DGAPA (UNAM) grant IN108207, from the CONACyT grants 46828-F and
61547, and from the ``Macroproyecto de Tecnolog\'\i as para la Universidad
de la Informaci\'on y la Computaci\'on'' (Secretar\'\i a de Desarrollo
Institucional de la UNAM).
\end{acknowledgements}


\end{document}